\documentclass[pra,aps,amsmath,amssymb,superscriptaddress,twocolumn,longbibliography]{revtex4-1}

\usepackage{graphicx,multirow}
\usepackage{color,outlines}
\usepackage[sc,osf]{mathpazo}
\usepackage{hyperref}
\usepackage{physics}
\usepackage{outlines}
\usepackage{float}
\begin{document} 
\title{ Probing quantum phase transition via quantum speed limit}
\author{M Suman}
\email{msuman.physics@gmail.com }
\author{S. Aravinda}
\email{aravinda@iittp.ac.in }    
\author{Ranjan Modak}
\email{ranjan@iittp.ac.in }
\affiliation{ Department of Physics, Indian Institute of Technology Tirupati, Tirupati, India~517619} 

%+++++++++ABSTRACT+++++++++++++++++++++++++++++++++
\begin{abstract}
Quantum speed limit (QSL) is the lower bound on the time required for a state to evolve to a desired final state under a given Hamiltonian evolution. 
Three well-known QSLs exist Mandelstam-Tamm (MT), Margolus-Levitin (ML), and dual ML (ML$^*$) bounds. We 
consider one-dimensional systems that undergoes delocalization-localization transition in the presence of quasiperiodic and linear potential. By performing sudden quenches across the phase boundary, we find that the exact dynamics get captured very well by QSLs.  We show that the MT bound is always tighter in the short time limit for any arbitrary state, while the optimal bound for the time of orthogonalization (time required to reach the orthogonal state) depends on the choice of the initial state. Further, for extreme quenches, we prove that the MT bound remains tighter for the time of orthogonalization, and it can qualitatively describe the non-analyticity in free energy for dynamical quantum phase transition (DQPT). We also demonstrate that the localization-delocalization transition point can be exactly identified from QSLs for two of the models of non-interacting fermions. Further, we also show that even for interacting systems, ergodic to many-body localization transition can also be efficiently predicted from QSLs, whose computation cost is much less than other diagnostic tools.

\end{abstract}
%+++++++++++++END OF ABSTRACT+++++++++++++++++++++

\maketitle

%=====================MAIN========================
\raggedbottom
\section{Introduction}
The quantum speed limit (QSL) is a fundamental limit in quantum mechanics on the rate at which any quantum system evolves under a dynamical process~\cite{anandan1990geometry,mandelstam1945uncertainty,okuyama2018quantum,shanahan2018quantum}.
It essentially provides a bound on the minimal time required to transport a
system from its initial state to a final state under unitary evolution. 
Apart from this theoretical interest in understanding the basic features of a quantum system, the quantum speed limit also remains an integral part of
recent advances in quantum engineering to investigate how quickly a transition can take place between distinguishable states~\cite{PhysRevA.102.042606,PhysRevA.106.L040401}. In other practical examples, such as to estimate the speed of quantum simulations
involving quantum information processing~\cite{PhysRevA.95.042314}, quantum computation, and quantum metrology~\cite{PhysRevLett.103.240501,lloyd.02,Maleki2023-cg}, identifying
decoherence time~\cite{PhysRevX.6.021031,PhysRevLett.118.140403,PhysRevLett.103.240501},  experimental measuring of the
environment-assisted speed-up~\cite{prl15}, even in the context of the  machine learning~\cite{ml}, QSL
plays a vital role.

The first QSL was
proposed by Mandelstam and Tamm (MT), who realized that fundamentally 
the time-energy uncertainty relation corresponds to the intrinsic timescale of unitary evolution in quantum mechanics~\cite{mandelstam1945uncertainty}. This relation restricts
the minimal time $\tau_{MT}$ for a unitary system to propagate between
two states that depend on its energy spread. On the other hand, time limit $\tau_{ML}$ was derived by
Margolus and Levitin (ML)~\cite{margolus1998maximum}, which depends on
the mean energy measured relative to the energy of the the ground state. Very recently, a QSL has been proposed which is referred to as dual ML bound (ML$^*$)~\cite{sagi.22}.
This bound is essentially equivalent to the ML bound in time-reversed dynamics. While    Levitin and Toffoli have unified the MT and the ML bound~\cite{levitin.09},  with the addition of ML$^*$ bound all three QSLs can be unified, and the lower bound for the  time required for arriving at an orthogonal
state (time of orthogonalization) $t_{\perp}$ can be written as,
\begin{equation}
    t_{\perp} \geq \tau_{QSL}=\text{max}\{\tau_{MT},\tau_{ML},\tau_{ML^{*}}\}.
    \label{main}
\end{equation}

The phase transition in many-body systems is a most illustrious phenomenon. In equilibrium many-body systems, it is witnessed as the non-analytical behaviour of its partition function at the thermodynamic limit~\cite{Fisher,mussardo2010statistical}. Motivated by this, the dynamical quantum phase transition (DQPT) is proposed as non-analytical behavior in the time expansion of the overlap amplitude of the time-evolved initial
quantum state~\cite{heyl2013dynamical,dqpt1,dqpt2,dqpt3}. 
The DQPT is characterized by the occurrence of zeros in the Loschmidt echo $\mathcal{L}$ at specific critical times $t^*_n$~\cite{heyl2013dynamical,le14,PhysRevB.95.060504}, 
where the Loschmidt echo is defined by $\mathcal{L}(t) = |\langle \psi_i|e^{-iH_ft}|\psi_i\rangle|^2$, and 
 $|\psi_i\rangle$ is the ground state of pre-quench Hamiltonian $H_i$, $H_f$ 
is the post-quench Hamiltonian.
 These zeros of $\mathcal{L}$  correspond to the 
non-analytic behaviors of the dynamical free energy $f(t) = -\frac{1}{N}\ln \mathcal{L}(t)$ at
those critical times~\cite{Heyl2018-uf}. In general, exact zeros of $\mathcal{L}$
 or non-analyticities of the dynamical free energy only occur as the system size $N$ approaches the thermodynamic limit, however, recent studies have shown
that they can be observed for finite systems as well~\cite{lefinite1,lefinite2}. The
critical time at which the first exact zeros of $\mathcal{L}$ occurs i.e. $t^*_1$, can be identified as the minimum time required for an initial state to reach an orthogonal state under the time evolution. From the definition, it is obvious that $t^*_1 \geq \tau_{QSL}$. 

The DQPT has been mostly observed in systems that go through an underlying equilibrium phase transition, and it is a generic feature of quantum
quenches across quantum critical points. It also has been observed in topological~\cite{top1,top2} and Floquet systems~\cite{floq1,floq2}. DQPT is also realized in experiments~\cite{exp1,exp2}.

%On the other hand, so far in most of the studies, QSLs have been investigated for few-level systems such as qubits~\cite{sagi.22}.  

Our main goal here is to investigate QSL for a Hamiltonian system that can be probed in experiments (e.g. in cold-atom experiments). 
Given the systems that show the signature of  DQPT have the unique features of having zeroes in the time evolution of $\mathcal{L}$, they become a natural choice to study QSLs for the time of orthogonalization. In this work, we focus on systems that display DQPT while quenching across the localization-delocalization transition point~\cite{dqpt_aa.17,modak.21,Faridfar_2023}. The dynamical regimes of applicability of particular QSL had been studied for few-level systems (qutrits)~\cite{sagi.22}, and it is important to investigate it to experimentally realizable physical many-level systems.

The aims of our study are two-fold: 1) to investigate which one of the QSLs can more efficiently capture short-time dynamics and the non-analytic nature of the dynamical free energy, 
%{\red thereby capturing the DQPT efficiently}
and 2) can the localization-delocalization transition point be identified using QSLs instead of studying the exact dynamics of the system?. We explore to answer these questions
by studying two one-dimensional non-interacting models. Aubry-Andre (AA) model~\cite{aubry1980analyticity}, and Wannier Stark model that supports localization transition~\cite{emin1987existence,RevModPhys.34.645}, and the one-dimensional interacting many-body Hamiltonian, which shows the transition from ergodic to Many-Body Localized (MBL) phase~\cite{abanin2019colloquium}.

The manuscript is organized as follows. In Sec. II, we discuss the preliminaries, which involves a discussion of models we study in this manuscript and also different QSLs.  Analytical results for extreme quenches are presented in Sec. III and followed by numerical results in Sec. IV. 
Finally, we conclude in Sec. V.

\section{Preliminaries}
%\section{Quantum speed limit}
Here we briefly discuss the models that have been investigated in this manuscript and also define different QSLs, which we have used extensively to compare with exact dynamics in the subsequent sections.  
%\section{Incommensurate lattice}
\paragraph*{Model:} We study a system that is described by the following 
Hamiltonian on a one-dimensional lattice, 
\begin{equation}
    H(J,\Delta) = -J\sum_{j=1}^N \left[c_{j+1}^\dagger c_j + h.c.\right] + \Delta\sum_{j=1}^N \epsilon_j n_j,
    \label{Hamiltonian}
\end{equation}
where $c_j$ and $c_j^{\dagger}$ is fermionic annihilation and creation operator respectively, and $n_j = c_j^\dagger c_j$ is the number operator. $J$ is the hopping strength, $\Delta$ is the strength of the on-site potential, in unit of $\hbar$. 
We consider two types of on-site potentials, 1) incommensurate potential $\epsilon_n=\cos(2\pi\alpha n)$ with $\alpha$ being an irrational number, and  2) linear potential $\epsilon_n=n$. The former is known as the AA model and shows localization transition at $\Delta=2J$~\cite{aubry1980analyticity}. In the case of the latter, 
any infinitesimal value of $\Delta$ 
is sufficient to localize all the states, the phenomena known as Wannier Stark localization~\cite{emin1987existence,RevModPhys.34.645}. 
Both of these Hamiltonians have been realized in the experiments~\cite{aa_exp1,aa_exp2,aa_exp3,stark_exp1,stark_exp2,Roati2008,Michel0001}. 
Motivated by the experimental setup, we mostly focus on a quench from localized to delocalized phase. Initially, we prepare our system as an eigenstate $|\psi_0\rangle$ of $H(\Delta_i)$ with $\Delta_i$ being very large, and keeping $J$ constant throughout the process.  Next, we quench it to the delocalized phase and study the time evolution of the Loschmidt echo  $\mathcal{L}(t)$. 
 At the end of our manuscript, we investigate the quench from the delocalized phase to the localized phase as well. For all our numerical results presented in the manuscript, we choose $\alpha=(\sqrt{5}-1)/2$ and open boundary condition (OBC), also $J = 1$.

We also test our prediction for the interacting many-body systems of the fermions, and the Hamiltonian is given by~\cite{PhysRevB.87.134202},
\begin{align}
    H_{MBL}(J,\Delta,V) = &-J\sum_{j=1}^N \left[c_{j+1}^\dagger c_j + h.c.\right] + \Delta\sum_{j=1}^N \epsilon_j n_j \nonumber\\
    &+ V\sum_{j=1}^N n_jn_{j+1},
    \label{MBLHamiltonian}
\end{align}
where $V$ is the nearest-neighbor interaction strength between the fermions, in unit of $\hbar$. 
In the absence of the on-site potential, i.e., $\Delta=0$, the Hamiltonian is Integrable, and the model is solvable within the Bethe-Ansatz framework~\cite{faddeev1996algebraic}.  However, for non-zero $\Delta$, there are no analytical solutions for this Hamiltonian, forcing us to use only numerical methodology. In our analysis, we focus on the quasi-random onsite potential, $\epsilon_j = \cos (2\pi\alpha j + \phi)$ with some phase factor $\phi$. This Hamiltonian shows ergodic to many-body localization phase transitions \cite{abanin2019colloquium} as one increases the strength of the on-site potential.

\paragraph*{Quantum speed limits:} Three QSLs bounds have been already introduced in the introduction section, where the consideration was that the time evolved state is orthogonal to the initial state $|\psi_0\rangle$. 
After a sudden quench is performed using a post-quench Hamiltonian 
 $H(\Delta_f)$, QSLs can be derived for any arbitrary state that is not necessarily orthogonal to $|\psi_0\rangle$. Three QSL bounds for any time evolved state $|\psi_T\rangle$ are given by,

\begin{equation*} 
    T_{MT} = \frac{1}{\Delta H} \arccos(|\langle\psi_0|\psi_T\rangle|), \nonumber
    \label{MT}
\end{equation*}

\begin{equation}
    T_{ML} = \frac{2}{\pi(\langle H\rangle - E_0)} (\arccos(|\langle\psi_0|\psi_T\rangle|))^2, 
    \label{ML}
\end{equation}

\begin{equation*}
    T_{ML^*} = \frac{2}{\pi(E_{max} - \langle H\rangle)} (\arccos(|\langle\psi_0|\psi_T\rangle|))^2\nonumber
    \label{ML*}
\end{equation*}

Average $\langle H(\Delta_f)\rangle$ and uncertainty $\Delta H=\sqrt{\langle H^2(\Delta_f)\rangle-\langle H(\Delta_f)\rangle^2}$ are  computed with respect to the initial state. $E_0$ and $E_{max}$ are the ground state and the highest excited state energy of the post-quench Hamiltonian $H(\Delta_f)$. In the subsequent sections, we compare the results obtained using exact dynamics and different QSLs, where we will use the above expressions extensively.

\section{Analytical Results for extreme quenches}

For a general quench  $\Delta_i\to \Delta_f$ across the phase boundary, computing QSLs analytically is not that straightforward. However, for extreme quenches i.e. $\Delta_i\to 0$ to $\Delta_f\to \infty$ (delocalized to localized) or $\Delta_i\to \infty$ to $\Delta_f\to 0$ (localized to delocalized), QSLs can be obtained analytically. 
Since we are quenching across the phase boundary, final state is expected to be orthogonal to initial state for AA and Wannier Stark model~\cite{Faridfar_2023,dqpt_aa.17}.
In this section, we focus on such cases and demonstrate that the MT bound is always tighter for extreme quenches, it can also show a clear signature of the dynamical quantum phase transition. 

Consider the Hamiltonian $H(J,\Delta=0)$, which is exactly solvable, and the eigenstate and eigenenergy are given by, 

\begin{equation}
    \begin{aligned}
        &|k\rangle = \frac{1}{N}\sum_{n=1}^N e^{ikn} c^\dagger_n |0\rangle\\
        &E_k = 2J\cos(ka)
    \end{aligned}
    \label{DelocalisedState}
\end{equation}

\noindent where, $k=\frac{2\pi(l-N/2)}{aN} \in\left(-\frac{\pi}{a},\frac{\pi}{a}\right]$ $(l=1,...,N)$ lies in first Brillouin zone. These eigenstates are delocalized plain waves. Similarly the Hamiltonian $H(J,\Delta \rightarrow \infty)$, has following eigenstate and eigenenergy,

\begin{equation}
    \begin{aligned}
        &|m\rangle = \sum_{n=1}^N\delta_{nm}c^\dagger_n|0\rangle\\
        &E_m = \Delta \epsilon_m.
    \end{aligned}
    \label{LocalisedState}
\end{equation}

\noindent These states are localized. Throughout this manuscript, we refer $|k\rangle$ as a delocalized state and $|m\rangle$ as a localized state in an extreme quench regime.  Also, for analytical calculation we keep $J$ constant, without mentioning in the further calculation.

\subsection{Quench from localized to delocalized phase}

\paragraph*{AA model:} First, we focus on the AA model and consider the case where the initial state is an eigenstate of the Hamiltonian $H(\Delta_i \rightarrow \infty)$, hence localized by definition. This state is quenched by the Hamiltonian $H(\Delta_f = 0)$. We set $J$ to be constant throughout the quench. We study which bound is tighter for such quench for the time of orthogonalization, and also compare the QSL bound with the results obtained from the exact dynamics.

To calculate ML and MT bound, we begin by calculating the expectation of the post-quench Hamiltonian with respect to the initial state.

\begin{equation}
    \langle H(\Delta_f=0)\rangle = \langle0|c_m H c_m^{\dagger}|0\rangle = 0
    \label{AA_Average}
\end{equation}

\noindent The expectation value is zero because diagonal elements of the Hamiltonian are zero. Similarly, the expectation value of the square of the Hamiltonian reads as, 

\begin{equation}
    \langle H^2(\Delta_f=0)\rangle = 2J^2.
\end{equation}

The details of the calculations are included in Appendix A. The above results are true for all $m$ except $m = 1$ and $m = N$. Hence, the uncertainty in energy of the post-quench Hamiltonian $\Delta H = \sqrt{\langle H^2 \rangle - \langle H\rangle^2} = \sqrt{2}J$. By knowing  the ground state energy $E_0 = -2J$ of the quench Hamiltonian, ML, and MT bound for the time of orthogonalization can be written as,

\begin{equation}
    \begin{aligned}
        &\tau_{ML}^{\infty\rightarrow0} = \frac{\pi}{2(\langle H \rangle - E_0)} = \frac{\pi}{4J}\\
        &\tau_{MT}^{\infty\rightarrow0} = \frac{\pi}{2\Delta H} = \frac{\pi}{2\sqrt{2}J}.
        \label{QSL_AA_12}
    \end{aligned}
\end{equation}

From the above equation, it is obvious that $\tau_{MT} > \tau_{ML}$. Hence for an extreme quench, the MT bound is always tighter than the  ML bound.

Next, to understand whether $\tau_{MT}$ is a good approximation to the orthogonalization time $t_\perp$ or not, we turn our attention towards deriving the results for the exact time-evolution of Loschmidt amplitude $\mathcal{G}(t)$, which reads as~\cite{dqpt_aa.17}, 
\begin{equation*}
    \begin{aligned}
        &\mathcal{G}(t) = \langle m| e^{-iH(\Delta_f)t} |m\rangle
        = \sum_{k=1}^N \langle m| ^{-iH(\Delta_f)t} |k\rangle\langle k|m\rangle\\
        &=\sum_{k=1}^N  e^{-2iJt\cos(ka)}|\langle m|k\rangle|^2 = \frac{1}{N}\sum_{k=1}^N e^{-2iJ\cos(ka)t},
    \end{aligned}
\end{equation*}
where $k$ is distributed over $-\pi/a$ to $\pi/a$. Hence, for a large system size $N\to \infty$, summation can be replaced by an integration,
\begin{equation*}
    \mathcal{G}(t) = \frac{a}{2\pi}\int_{-\frac{\pi}{a}}^{\frac{\pi}{a}} e^{-2iJ\cos(ka)t} dk = \mathcal{J}_0(2Jt),
\end{equation*}
where $\mathcal{J}_0$ is the zeroth order Bessel function. Its zeros occur at,
\begin{equation}
    t^\star_\alpha = \frac{x_\alpha}{2J},
    \label{DQPTLocaltodelocal}
\end{equation}
where $x_\alpha$ are roots of the Bessel function. From Eq.~\eqref{DQPTLocaltodelocal} and \eqref{QSL_AA_12}, we find that the time of orthogonalization obtained from the QSL bound as well as from the exact dynamics, both are independent of $\Delta_f$. The exact first zero of $\mathcal{L}(t)$ occurs at $t^*_1 = 1.2024/J \geq \tau_{MT}= 1.1072/J$. Hence, MT bound serves as a very good approximation for the exact result.

\paragraph*{Wannier Stark model:} For the Wannier Stark model, both ML bound and MT bound give the same results as one obtained for the AA model, as the eigen state of pre-quench Hamiltonian and post-quench  Hamiltonian are the same for both models for extreme quenches. Also, Dual ML bound for the Wannier Stark model is not defined, as this model do not have upper spectral bound to its energy. QSL bounds for the Wannier Stark model are distinguished by a tilde symbol that reads as, 

\begin{equation}
\begin{aligned}
    &\tilde{\tau}_{ML}^{\infty\rightarrow0} = \frac{\pi}{2(\langle H \rangle - E_0)} = \frac{\pi}{4J}\\
    &\tilde{\tau}_{MT}^{\infty\rightarrow0} = \frac{\pi}{2\Delta H} = \frac{\pi}{2\sqrt{2}J}
\end{aligned}
\label{QSL_Stark_12}
\end{equation}

The orthogonalization time obtained from the exact dynamics is also the same as before i.e. $t^*_{\alpha} = x_\alpha/2J$.

\subsection{Quench from delocalized to localized phase}
\paragraph*{AA model:} Here, we consider the initial state as the eigenstate of the Hamiltonian $H(\Delta_i=0)$ and post-quench Hamiltonian is AA Hamiltonian $H(\Delta_f \rightarrow\infty)$.

We compute $\langle k| H| k\rangle$ and $\langle k| H^2|k\rangle$ for the post-quench Hamiltonian $H(\Delta_f \rightarrow \infty)$. In the limit  $N\to \infty$, they are given by, 
\begin{equation}
\begin{aligned}
    &\langle H(\Delta_f\rightarrow\infty)\rangle = \frac{1}{N}\sum_{n,p=1}^N e^{-ink}\langle0|c_{n}H c_{p}^\dagger|0\rangle e^{ipk}\\
    &= \frac{\Delta_f}{N} \sum_{n=1}^N \cos(2\pi\alpha n) \approx 0,
    \label{Average}
\end{aligned}
\end{equation}
and 
\begin{equation}
    \langle H^2(\Delta_f\rightarrow\infty)\rangle = \frac{\Delta_f^2}{N} \sum_{n=1}^N \cos^2(2\pi\alpha n) \approx \frac{\Delta_f^2}{2}.
    \label{AverageSquare}
\end{equation}

The energy uncertainty  $\Delta H = \Delta_f/\sqrt{2}$ and the ground state energy of the post-quench Hamiltonian is $E_0 = -\Delta_f$. The QSL bound for the time of orthogonalization can be written as,

\begin{equation}
    \begin{aligned}
        &\tau_{ML}^{0\rightarrow\infty} = \frac{\pi}{2\Delta_f}\\
        &\tau_{MT}^{0\rightarrow\infty} = \frac{\sqrt{2}\pi}{2\Delta_f}.
    \end{aligned}
    \label{QSL_AA_21}
\end{equation}

It automatically implies that $\tau_{MT} > \tau_{ML}$.
Once again, to find out how tight the QSL bound is in comparison to the exact dynamics, we turn our attention to the exact results, and the time evolution of the exact Loschmidt amplitude can be written as~\cite{dqpt_aa.17},
\begin{equation*}
\begin{aligned}
    &\mathcal{G}(t) = \langle k|e^{-iH(\Delta_f)t}|k\rangle
    =\sum_{m=1}^N \langle k|e^{-iH(\Delta_f)t}|m\rangle\langle m|k\rangle\\
    & = \sum_{m=1}^N e^{-i\Delta_f\cos(2\pi\alpha m)t}|\langle m|k\rangle|^2.
\end{aligned}
\end{equation*}
Now from Eq.~\eqref{DelocalisedState} and \eqref{LocalisedState}, $|\langle m|k\rangle|^2 = 1/N$. Hence,
\begin{equation*}
    \mathcal{G}(t) = \frac{1}{N}\sum_{m=1}^N e^{-i\Delta_f\cos(2\pi\alpha m)t}.
\end{equation*}

For an irrational number $\alpha$, the phase $2\pi\alpha m$ is distributed between $-\pi$ and $\pi$. In $N\to \infty$ limit, the summation can be converted into an integral, and 

\begin{equation*}
    \mathcal{G}(t) \approx \frac{1}{2\pi}\int_{-\pi}^{\pi} e^{-i\Delta_f\cos(\theta)t} d\theta = \mathcal{J}_0(\Delta_ft).
\end{equation*}

Zeros of the Bessel function $\mathcal{J}_0(\Delta_ft)$ correspond to the zeros of the Loschmidt amplitude, they are given by,

\begin{equation}
    t^*_\alpha = \frac{x_\alpha}{\Delta_f}
\end{equation}

The first zero is $t^*_1 = 2.4048/\Delta_f > \tau_{MT} = 2.2214/\Delta_f$, and interestingly both the exact result and MT bound are inversely proportional to $\Delta_f$, and they are reasonably close.

\paragraph*{Wannier Stark model:}
Unlike the previous scenario, QSL bounds for the extreme quench from the delocalized phase to the localized phase give rise to different results for the Wannier Stark model compared to the AA model. For the Wannier Stark model, 
$\langle H(\Delta_f\rightarrow\infty)\rangle$ and $\langle H^2(\Delta_f\rightarrow\infty)\rangle$ are given by,

\begin{equation}
\begin{aligned}
    &\langle H(\Delta_f\rightarrow\infty)\rangle = \frac{1}{N}\sum_{n,p=1}^N e^{-ink}\langle0|c_{n}H c_{p}^\dagger|0\rangle e^{ipk}\\
    &= \frac{\Delta_f}{N} \sum_{n=1}^N n = \Delta_f \frac{(N+1)}{2}\\
    &\langle H^2(\Delta_f\rightarrow\infty)\rangle = \frac{\Delta_f^2}{N} \sum_{n=1}^N n^2 = \Delta_f^2 \frac{(N+1)(2N+1)}{6}
\end{aligned}
\end{equation}

The uncertainty , $\Delta H = \Delta_f \sqrt{\frac{N^2 - 1}{2}}$. Noting that the ground state energy of $H(\Delta_f\rightarrow \infty)$ is zero, the QSL bound can be written as,

\begin{equation}
    \begin{aligned}
        &\Tilde{\tau}_{ML}^{0\rightarrow\infty} = \frac{\pi}{\Delta_f(N+1)}\\
        &\Tilde{\tau}_{MT}^{0\rightarrow\infty} = \frac{\sqrt{3}\pi}{\Delta_f\sqrt{N^2-1}}
    \end{aligned}
    \label{QSL_Stark_21}
\end{equation}

Once again,  it is obvious that $\Tilde{\tau}_{MT} > \Tilde{\tau}_{ML}$, which automatically makes MT a tighter bound. Now to compare $\Tilde{\tau}_{MT}$ with the zeros of the exact Loschmidt amplitude,  we compute $\mathcal{G}(t)$, i.e.

\begin{equation*}
\begin{aligned}
    &\mathcal{G}(t) = \langle k|e^{-iH(\Delta_f)t}|k\rangle
    =\sum_{m=1}^N \langle k|e^{-iH(\Delta_f)t}|m\rangle\langle m|k\rangle\\
    & = \sum_{m=1}^N e^{-i\Delta_f m t}|\langle m|k\rangle|^2 = \frac{1}{N}\sum_{m=1}^N e^{-im\Delta_f t}.
\end{aligned}
\end{equation*}

With some simplifications, one can end up with $\mathcal{L}(t) = |\sin(x)/x|^2$, where $x = \Delta_ftN/2$. $\mathcal{L}(t)$ has zeros at $t^*_n = \frac{2\pi n}{\Delta_f N}$. Hence, for a large but finite value of $N$, the first zero is given by, $t^*_1 = \frac{2\pi}{\Delta_f N} > \tilde{\tau}_{MT}$. The important thing to note here is that the orthogonalization time is inversely proportional to the system size $N$, whereas for the AA model, they were independent of $N$. In the thermodynamic limit $N\to \infty$, both $t^*_1$, $\tilde{\tau}_{MT} \to 0$.

\subsection{Detection of transition point}
Our main finding from the previous section was, for the extreme quenches $\tau_{QSL}=\tau_{MT}$, i.e. 
MT bound is always tighter for the time of orthogonalization, and we also obtain the exact expression for 
$\tau_{MT}^{0\rightarrow\infty}$ and $\tau_{MT}^{\infty\rightarrow0}$. For the AA model, interestingly, it turns out that $\tau_{MT}^{0\rightarrow\infty}=\tau_{MT}^{\infty\rightarrow0}$ when $\Delta_f=2J$, which is also the transition point for this model.

If one draws a conclusion from the AA model that at the transition point $\tau_{QSL}$ should be the same for two opposite extreme quenches, for the Wannier Stark model, once again, one would expect that   $\tilde{\tau}_{MT}^{0\rightarrow\infty}=\tilde{\tau}_{MT}^{\infty\rightarrow0}$ at the boundary of the phase-transition. It implies  
 $\Delta_f = \sqrt{\frac{24}{N^2 - 1}}J$, and the transition point for the Wannier Stark model should scale to zero as $N^{-1}$ in the $N\to \infty$ limit. Indeed, it is well-known that the transition point approaches zero for the Wannier Stark model in the thermodynamic limit~\cite{Faridfar_2023}. 

\section{Numerical Results}
Earlier, we focused on the extreme quenches. Here, we show the numerical results for 
$\Delta_i$ and $\Delta_f$ both are non-zero but finite. First, we discuss the quench from the localized to the delocalized phase. Then we show the results corresponding to the delocalized to localized quench. Finally, we demonstrate how the $\tau_{QSL}$ can be used to detect the transition point for AA and the Wannier Stark Hamiltonian, along with many body Hamiltonian. Also, we would like to remind that for all numerical results, we took $J=1$.

\subsection{Quench from localized phase to delocalized phase}

\begin{figure}[H]
    \centering
    \includegraphics[width=0.9\linewidth]{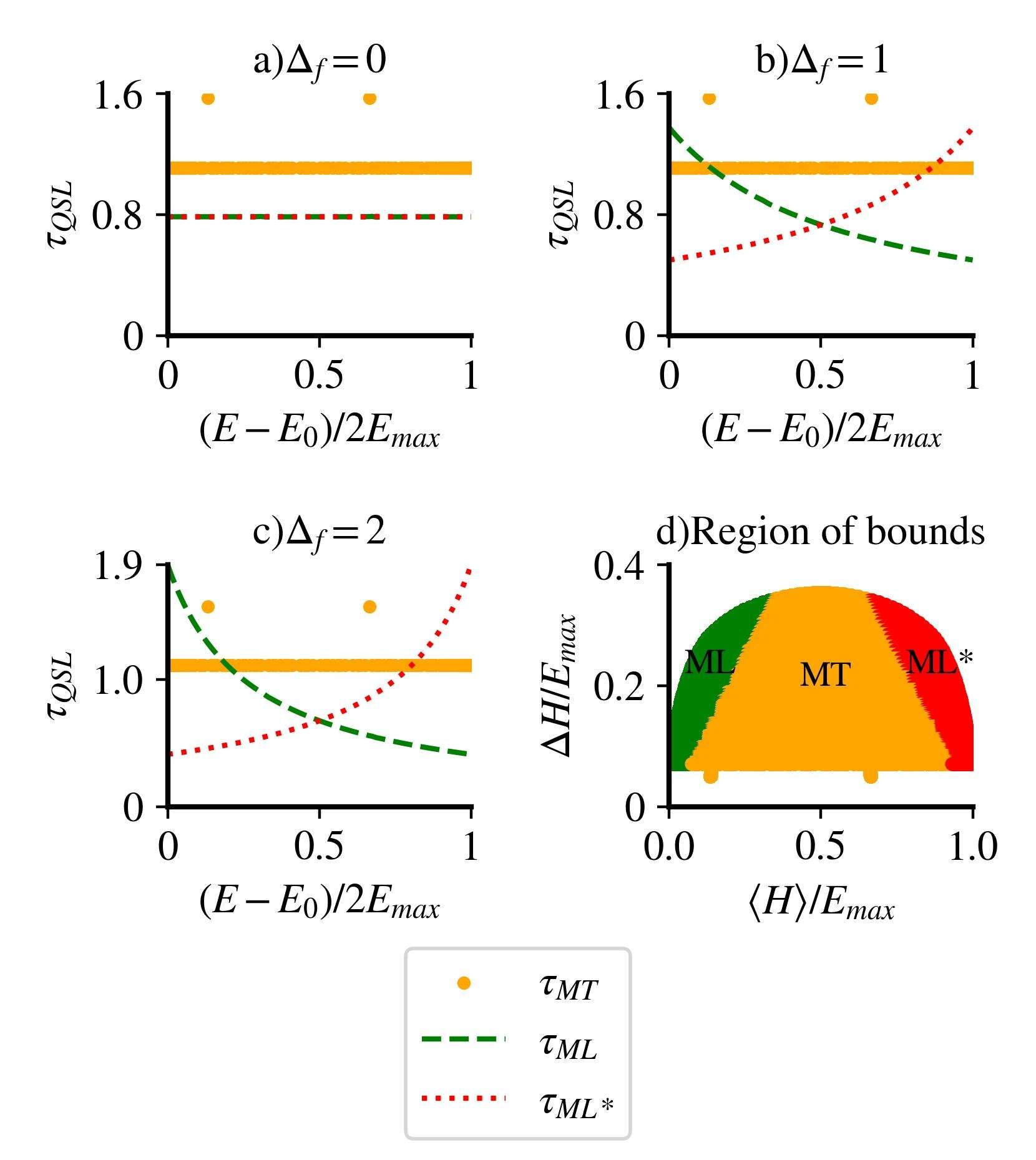}
    \caption{Region of best bounds. In the figures a,b, and c, each bound as a function of energy are plotted for different $\Delta_f$ values. In this figure, we choose $\Delta_i=1000$ and system size, $N=100$. In the figure d re-scaled $\Delta H$ vs $\langle H\rangle$ is plotted. Hamiltonian's ground state is already set to zero ($H = H(\Delta_i) - E_0$). The initial states are taken as eigenstates of $H(\Delta_i=1000)$. The quench Hamiltonian $H(\Delta_f)$ is taken for different $\Delta_f$, varying from $0$ to $10$. }
    \label{fig:1}
\end{figure}

We focus on the AA model and consider the situation where the initial state is still highly localized eigenstate of $H(\Delta_i>>2)$, and  $\Delta_f < 2$.  It has been shown in the previous section that for $\Delta_f=0$, the MT bound corresponding to the orthogonal time is tighter for any eigenstates of $H(\Delta_i>>2)$. 
From Fig.~\ref{fig:1}, it is clear that for $\Delta_f\neq 0$, the region of best bounds depends on the initial state i.e. energy eigenstate of $H(\Delta_i)$. Figure.~\ref{fig:1} [(b)-(c)] shows, for the high energy and low energy states, ML$^*$ and ML bounds are tighter respectively, whereas for the intermediate energy states MT bound remains tighter. As expected, with decreasing $\Delta_f$, more and more states start respecting MT bound, and in the limit $\Delta_f\to 0$, MT becomes tighter for all the states.

When this bound was derived, we excluded the state corresponding to the first and last lattice points. For the first and last lattice points the MT bound can be written as, $\tau_{MT} = \pi/2J$ (see Appendix A). Bound at these points are always greater than the other points. Hence in the figure \ref{fig:1}, we can see those two points of MT bound are always higher than the rest of the values. Note that this is simply the boundary effect that will go away if we use the periodic boundary condition (PBC) instead of open boundary condition (OBC). 

%-------------------------FIRST FIGURE--------------------------------
%----------------------------------------------------------------------
The results shown in Fig.~\ref{fig:1} can be well understood by investigating energy expectation value and the uncertainty in energy for the post-quench Hamiltonian. 
Figure.~ \ref{fig:1} (d) gives an idea about regimes, where which bound becomes tighter (note that we have subtracted the ground state energy $E_0$ from the energy expectation $\langle H\rangle$).  The states for which $\Delta H \geq \langle H\rangle$ correspond to the green-shaded ML regime. Similarly, states with $\Delta H \geq E_{max} - \langle H\rangle$ belong to the red shaded ML$^*$ bound regime. The orange-shaded region corresponds to the MT bound. 
The upper limit of all these bounds is further constrained by Popoviciu’s inequality~\cite{popoviciu1935equations,sagi.22}), i.e., 
\begin{equation}
    \Delta H \leq \sqrt{\langle H\rangle(E_{max} - \langle H\rangle)}
\end{equation}
From this inequality, $\Delta H$ becomes maximum at $\langle H \rangle = E_{max}/2$. That is $\Delta H = E_{max}/2$. 
At the intersection of $\Delta H = \langle H\rangle$ and $\Delta H = E_{max} - \langle H \rangle$ lines, all three bounds will coincide~\cite{sagi.22}. For our Hamiltonian $H$, the $\Delta H$ never saturates the Popoviciu's inequality. Hence all three bounds never coincide with each other.

\begin{figure}[H]
    \centering
    \includegraphics[width=0.9\linewidth]{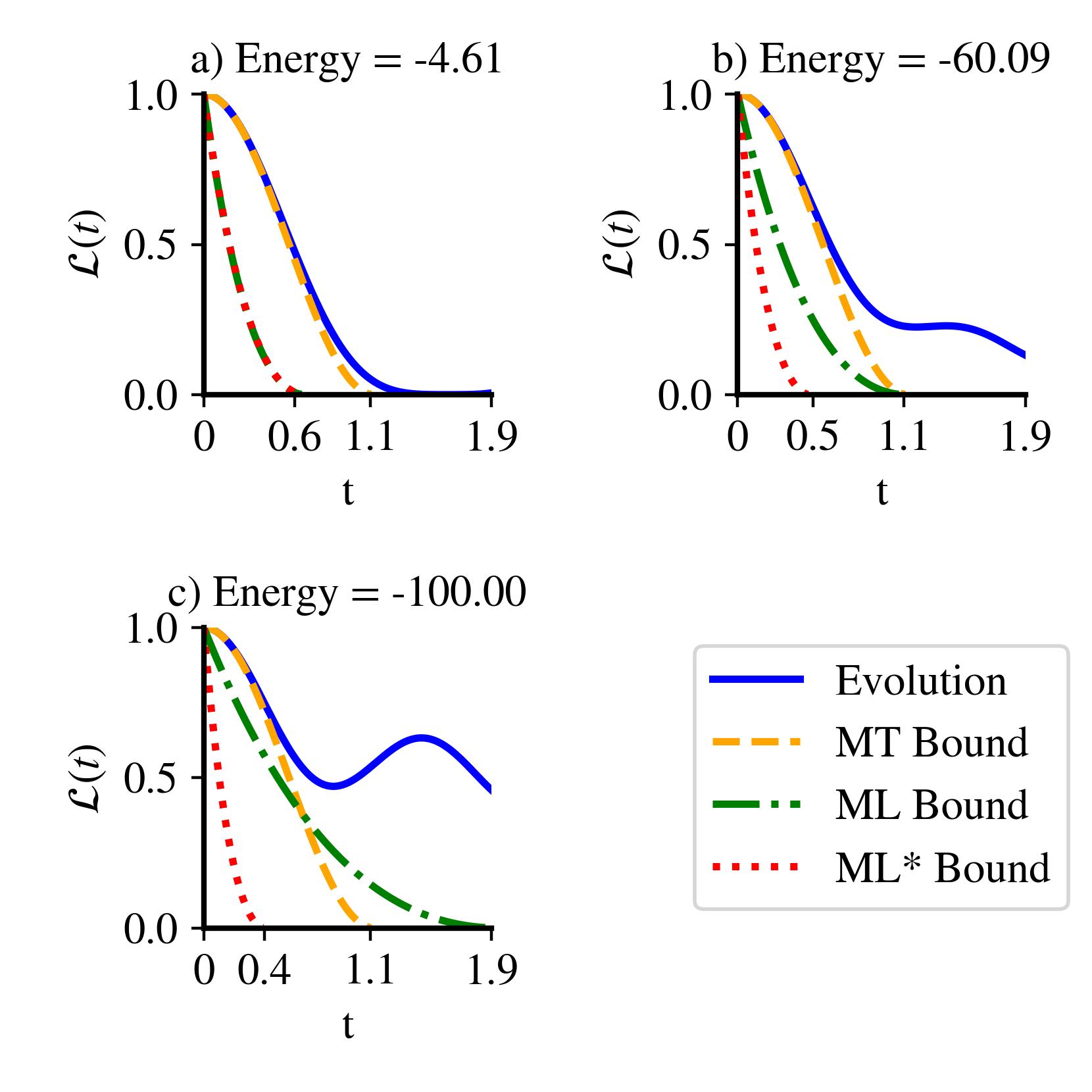}
    \caption{State evolution of the system in different dynamical regimes. From Eq.~ \eqref{ML}, Loschmidt echo is plotted as a function of time as well as for exact state evolution. Initial Hamiltonian is taken to be $H(\Delta_i=100)$ and quench Hamiltonian is taken as $H(\Delta_f=1.5)$. System size is, $N = 100$. In figure a, MT bound dominates and, the orthogonal time of MT bound and exact time evolution is quite close. In figure b, time of orthogonalization of exact evolution is at $t_\perp =  5.4$, in this case ML and MT both coincides. In figure c, time of orthogonalisation of exact evolution is $t_\perp = 33.9$, ML bound dominates in these energy state.}
    \label{fig:2}
\end{figure}

While so far we mostly focused on the bound corresponding to the time of orthogonalization $\tau_{QSL}$, here we redirect our attention to a time window  $[0, T]$, where $T< \tau_{QSL}$.
Figure.~\ref{fig:2} shows that not only MT bound is tighter for the mid-spectrum states, but the differences between the exact dynamics and the bound are also quite small. On the other hand, for low-energy states (and also for high-energy states)  the bound is not that close to the exact result. 
Moreover, in Fig.~\ref{fig:2} (c) we study a particular case, where the initial state and the post-quench Hamiltonian are chosen in such a way that for some time MT bound is tighter, followed by a crossover and finally ML bound becomes tighter. 

%------------------------SECOND FIGURE---------------------------------
%-----------------------------------------------------------------------

The fact that in the initial time, MT bound is tighter, can be well understood from 
the Eq.\eqref{ML}.  Inverting the Eq.\eqref{ML}, the Loschmidt echo corresponds to the bounds can be calculated as,

\begin{equation}
    \begin{aligned}
        &\mathcal{L}(t_{ML}) = \cos^2\left(\sqrt{\frac{\pi\langle H\rangle        t_{ML}}{2}}\right)\\
        &\mathcal{L}(t_{MT})  = \cos^2(\Delta Ht_{MT})
    \end{aligned}
    \label{crossover}
\end{equation}
At the crossover point both the bounds are equal, i.e.  $\Delta H T_{cross} = \sqrt{\frac{\pi\langle H\rangle T_{cross}}{2}}$ ;($t_{ML} = t_{MT} = T_{cross}$).  Hence, one gets the crossover time as, 
\begin{equation}
    T_{cross} = \frac{\pi\langle H\rangle}{2(\Delta H)^2}
\end{equation}
Also, the crossover takes place only if, $\Delta H\geq \langle H\rangle$. Including this additional constrain and from Eq.~\eqref{crossover}, it can be seen that $\mathcal{L}(t_{ML}) < \mathcal{L}(t_{MT})$ $\forall$  $t< T_{cross}$. This also has been observed in Fig.~\ref{fig:2}. 

\begin{figure}[H]
    \centering
    \includegraphics[width=0.9\linewidth]{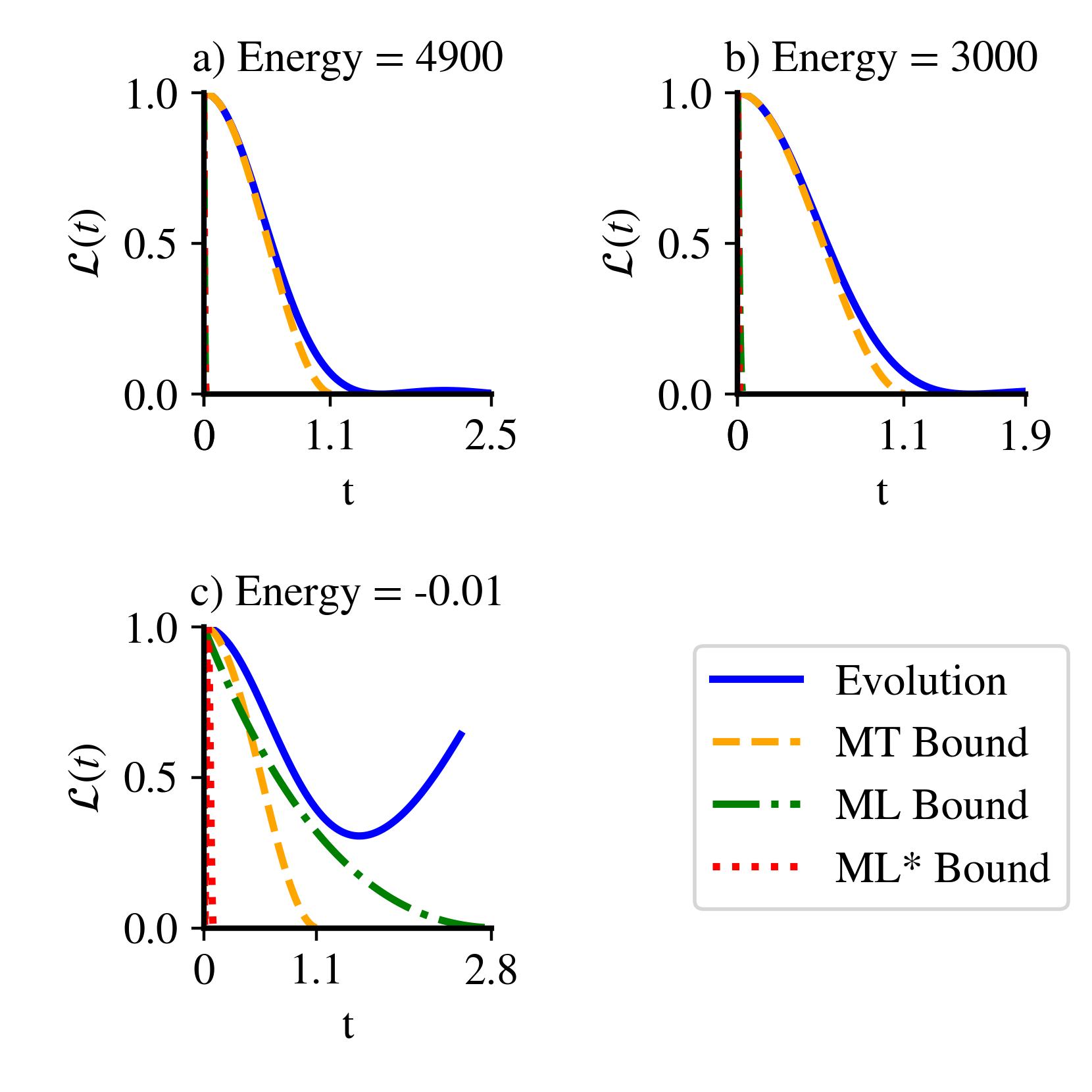}
    \caption{State evolution of the system for different energy regime for the Wannier Stark model. Analogous to the figure \ref{fig:2}, we plot Loschmidt echo from equation \eqref{ML} and from numerical simulation. Initial state is taken to be eigenstate of the Hamiltonian $H(\Delta_i = 100)$ and this state is quenched by the Hamiltonian $H(\Delta_f = 1.5)$. System size is $N=100$.}
    \label{fig:boundVsMosaic}
\end{figure}

Next, we repeat our argument for the Wannier Stark model. As expected, QSL in the Wannier Stark model shows behavior similar to that of the AA model. MT bound dominates in mid-spectrum, while ML bound dominates for low energy states. Even the crossover is observed in figure \ref{fig:boundVsMosaic}, which was also observed in the case of the AA model.

%=========================END OF SUBSECTION============================

\subsection{Quench from delocalized phase to localized phase}

Here we concentrate on the quench from the delocalized to the localized phase. First, we consider the AA model. Similar to the previous section results, once again, we find that  $\tau_{QSL}$ for low and high energy states are obeyed by the ML and ML$^*$ bound respectively, whereas the mid-spectrum states are MT. As we tend towards the extreme quench regime by increasing $\Delta_f$, the MT bound starts dominating over the entire spectrum. Given these results are very similar to the one demonstrated in Fig.~\ref{fig:1}, we do not show them explicitly in this section anymore. We choose to analyze in detail how well the bound depicts the exact evolution and the zeros in Loschmidt echo while quenching across the phase boundary. 
\begin{figure}[H]
    \centering
    \includegraphics[width=0.9\linewidth]{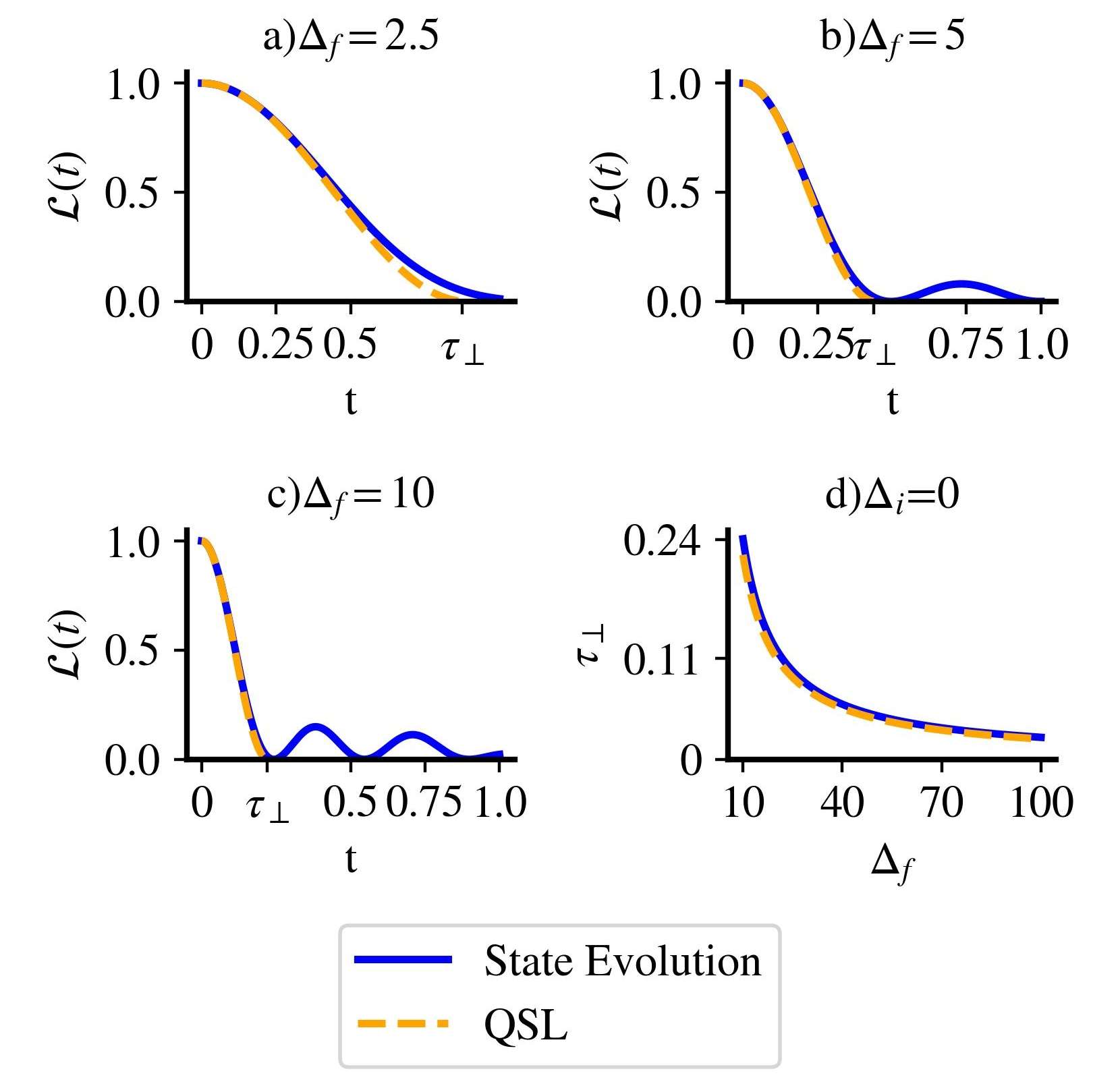}
    \caption{Comparison of bound with exact results for incommensurate lattice. We used equation \eqref{ML} to obtain Loschmidt echo}. Plot a,b, and c show consistency of QSL with change in $\Delta_f$. QSL is maximum of all three bounds. In the figure, $\Delta_i = 0$ and system size $N = 100$.  Figure d represents the first zero in Loschmidt echo and QSL at orthogonal time for different $\Delta_f$ values.
    \label{fig:3}
\end{figure}
%are the signature of Dynamical Quantum Phase Transition (DQPT) in both these models (\cite{dqpt_aa.17},\cite{Faridfar_2023}). 
In the Fig.~\ref{fig:3}, we show the results for $\Delta_f > 2$ for the mid-spectrum states. It can be seen that the MT bound matches with the exact evolution remarkably well until some time and then it starts to deviate from the exact evolution. The orthogonalization time of the bound is still quite close to the exact orthogonal time $t_{\perp}$.

\begin{figure}[H]
    \centering
    \includegraphics[width=0.9\linewidth]{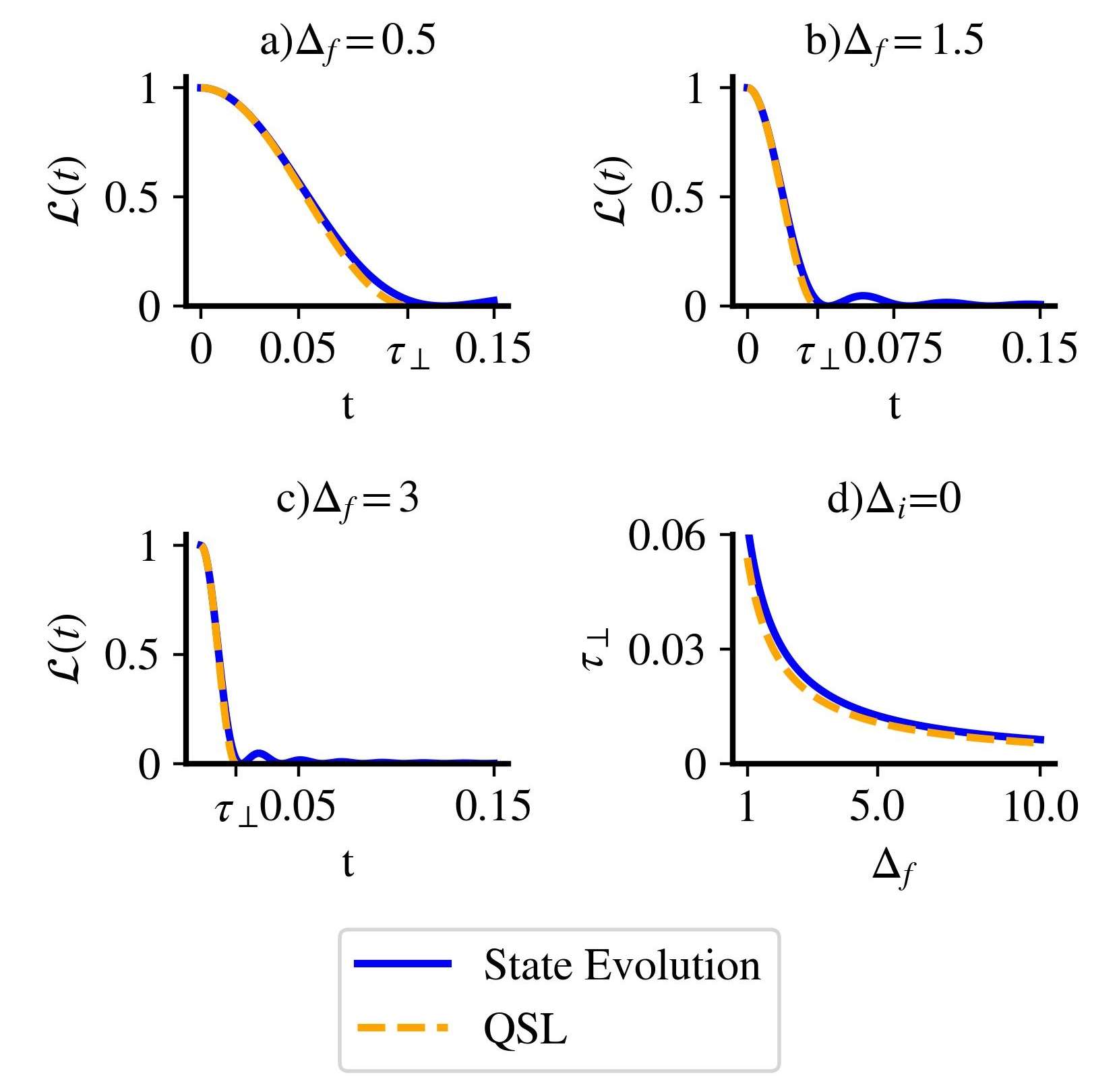}
    \caption{Comparison of bound with exact results for the Wannier Stark model. Plot a,b, and c show a comparison of QSL bound with the exact dynamics for different $\Delta_f$. In the figure $\Delta_i = 0$ and system size $N=100$. Figure d represents the first zero in Loschmidt echo vs $\Delta_f$. We have used equation \eqref{ML} to obtain $\mathcal{L}(t)$.}
    \label{fig:4}
\end{figure}

 Figure.~ \ref{fig:3} (d) also shows that the orthogonalization times obtained from both the bound and exact dynamics are inversely proportional to $\Delta_f$. The Wannier Stark model also shows the same behavior. In the case of the Wannier Stark model, all states are localized for $\Delta_f > 0$. In Fig.~\ref{fig:4}, one can see that the Loschmidt echo reaches zero much faster compared to the AA model. Once again like the AA model, the MT bound matches with the exact evolution remarkably. The orthogonalization time of the bound is reasonably close to the exact orthogonal time, both of them scale as $\Delta^{-1}_f$. This result confirms that indeed the QSL can mimic the exact dynamics of $\mathcal{L}(t)$ reasonably well for both the models. 

%-------------END OF SUBSECTION--------------------

\subsection{Detection of transition point from QSL \label{sec_IVD}}

\begin{figure}[H]
    \centering
    \includegraphics[width=0.9\linewidth]{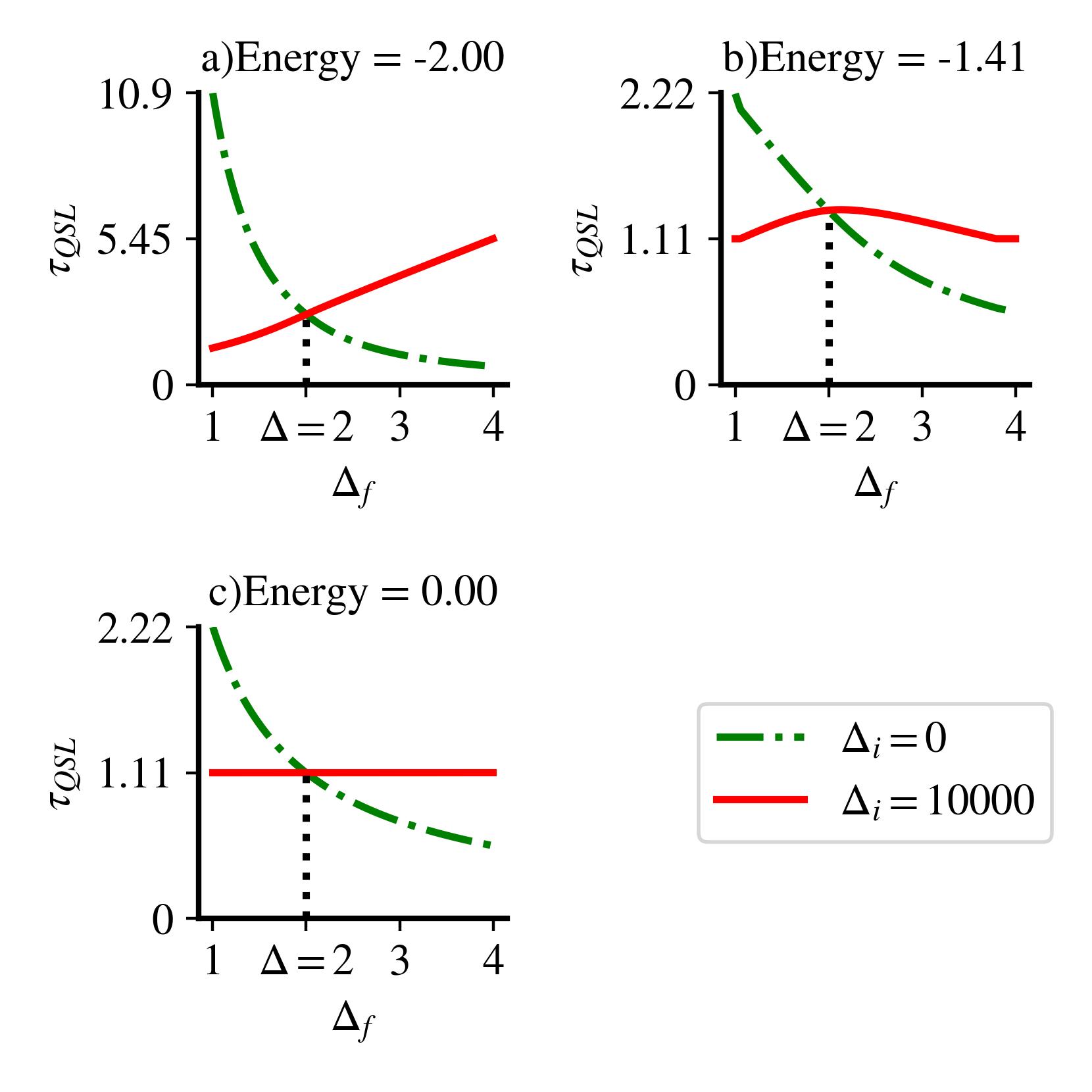}
    \caption{Plot of $\tau_{QSL}$ vs $\Delta_f$ for incommensurate lattice for different energy eigen kets. Energy labeled are Energy of eigen kets of the Hamiltonian $H(\Delta_i = 0)$. $\tau_{QSL}$ is the orthogonalization time of maximum of all three bounds(\ref{ML}). System size is taken to be $N = 1000$.}
    \label{fig:5}
\end{figure}

Here the goal is to identify the transition point using QSL. Motivated by the analytical results for the extreme quench, we study the variation of $\tau_{QSL}$ (see Eq.~\eqref{main}) with $\Delta_f$. We consider two situations, 1) $\Delta_i=0$ and 2) $\Delta_i>>2$. We find that in a $\tau_{QSL}$ vs $\Delta_f$ plot for the AA model (figure \ref{fig:5}), two curves (one corresponds to  $\Delta_i=0$ and another $\Delta_i=10000$) intersect at $\Delta_f=2$ for all initial states. Note that $\Delta=2$ also corresponds to the localization-delocalization transition point of the AA model.  
This result does not depend on $N$ as long $N$ is reasonably large.

\begin{figure}[H]
    \centering
    \includegraphics[width=0.9\linewidth]{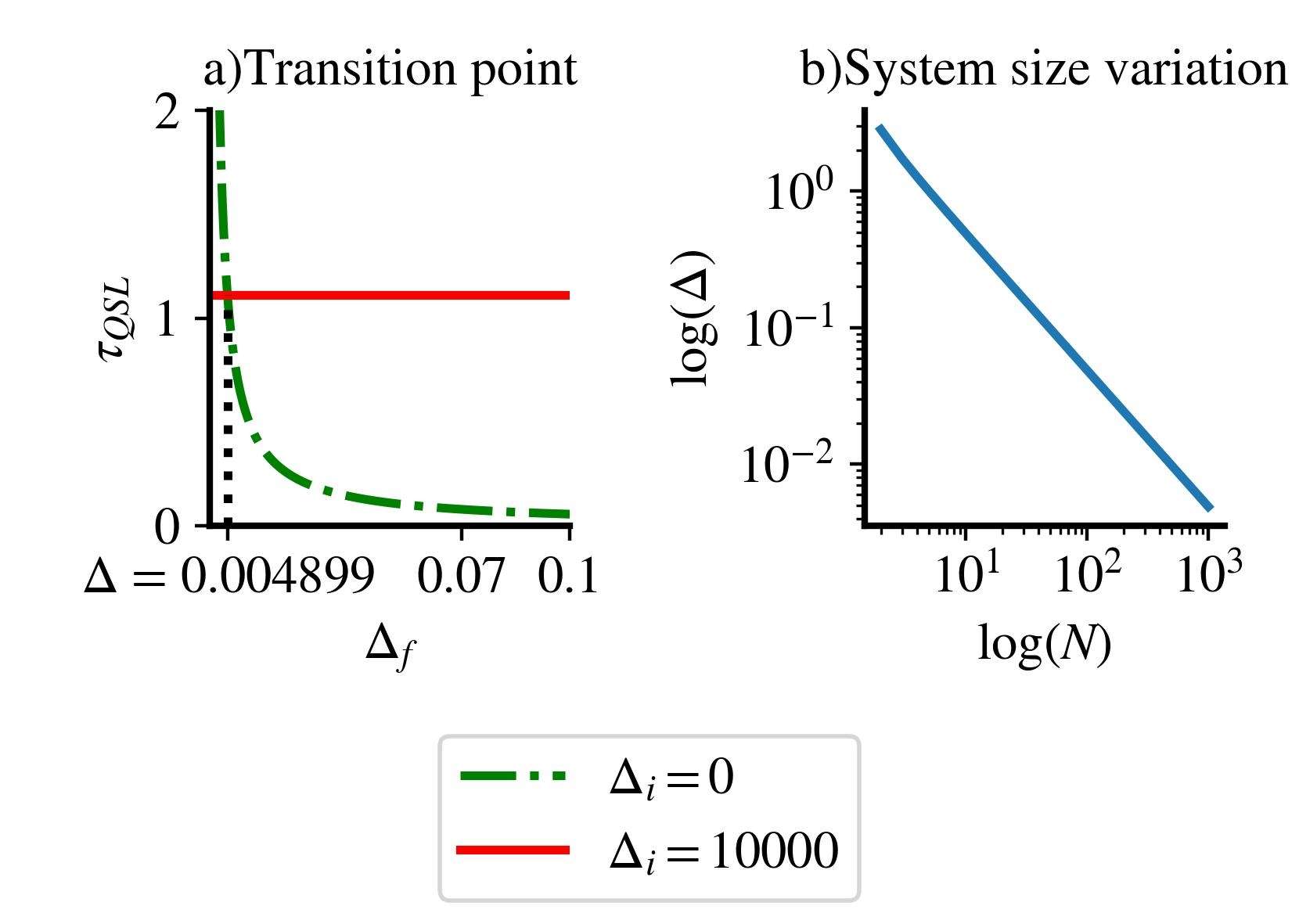}
    \caption{Transition point for the Wannier Stark model. Figure a represents the transition point for system size $N=1000$. Figure b represents loglog plot of the variation of transition point with system size $N$.}
    \label{fig:6}
\end{figure}

In contrast to the AA model, for the Wannier Stark Hamiltonian, in the $N\to \infty$ limit, any infinitesimal $\Delta$ is sufficient to localize all eigenstates.  However, for finite $N$, one can have a 
localization-delocalization phase transition for finite $\Delta$. By following the same procedure as before, we plot (see Fig.~\ref{fig:6} a)  $\tau_{QSL}$ vs $\Delta_f$, and two curves (one corresponds to  $\Delta_i=0$ and another $\Delta_i=10000$) intersect at $\Delta_f=0.004899$ for $N=1000$.

 This transition point depends on the system size $N$. In Fig.~ \ref{fig:6} b, we plot the phase transition point as a function of system size. From the figure, it is apparent that as system size increases, the transition point heads towards zero as $N^{-1}$. Our finding is consistent with the results obtained from Normalized participation ratio (NPR) calculations as well (see Appendix B). 

\subsection{Detection of MBL transition point from QSL}
Finally, we also considered the interacting system to test our prediction to its full potential. We focus on the Hamiltonian $H_{MBL}$~\eqref{MBLHamiltonian}. It is well-established that for a fixed interaction strength $V$, as we increase the strength of the incommensurate potential $\Delta$, the system undergoes ergodic to many-body localization transition. As we already discussed earlier, in the absence of interaction, the delocalization-localization transition point corresponds to $\Delta=2J$,  but with the increase of interaction strength, one needs a larger value of $\Delta$ to see the transition~\cite{PhysRevB.87.134202}.  Previously, in Sec.~\ref{sec_IVD}, we have demonstrated that the QSL can successfully predict $\Delta=2J$ transition point for a non-interacting system.  Now, the question arises whether the QSL can also capture the MBL transition point or not.

\begin{figure}[H]
    \centering
    \includegraphics[width=0.9\linewidth]{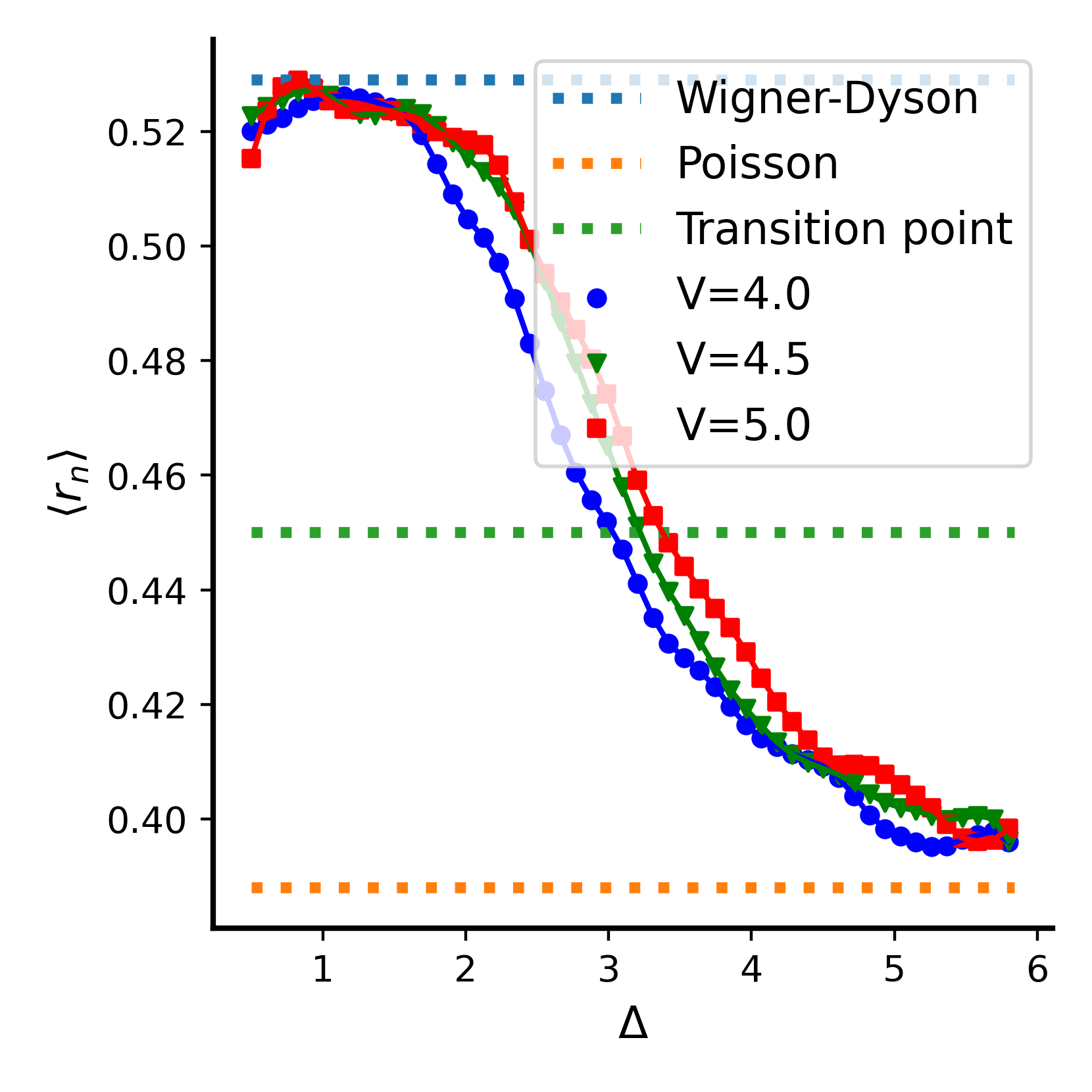}
    \caption{Plot of average level spacing characteristics vs. onsite potential $\Delta$. Lattice size is taken to be $N=12$ with half filling.}
    \label{fig:LevelSpacing}
\end{figure}

First, we use the level spacing characteristics to detect the MBL transitions. 
Level statistics of the many-body Hamiltonian varies from Wigner-Dyson statistics (ergodic phase) to Poissonian statistics (Many-body localized phase)\cite{PhysRevB.87.134202}.
If the energies of Hamiltonian are written in ascending order, $E_1, E_2,..., E_n$, the gap between successive energy is given as $\delta_n = E_{n+1} - E_n$. Then, the correlation between successive gaps in the spectrum can be written as
\begin{equation}
    r_n = \frac{min(\delta_n,\delta_{n+1})}{max(\delta_n,\delta_{n+1})}.
\end{equation}

We are mainly interested in the average correlation $\langle r_n \rangle$. For Poissonian statistics, we have $\langle r_n \rangle \approx 0.386$, and for Wigner-Dyson statistics, $\langle r_n \rangle \approx 0.5295$. Thus, when we vary $\Delta$ in the Hamiltonian $H_{MBL}$ (for a fixed $V$), we expect the value of  $r$ parameter to decrease from $0.529$ (Wigner-Dyson statistics) to $0.386$  (Poissonian statistics) with increasing $\Delta$ as we approach the ergodic to MBL phase. 
From figure \ref{fig:LevelSpacing}, we see that for $V=4, 4.5, 5$, after some critical value of $\Delta_c$, $\langle r_n \rangle$ reaches close to $0.529$, and then approaches  towards $0.386$ with increasing $\Delta$ for $N=12$. We like to emphasize that, though,  one would expect any tiny amount of $\Delta$ to ensure Wigner-Dyson statistics, given the Hamiltonian $H_{MBL}$ is non-integrable; however, it has been shown that for the finite-sized system,  one needs a critical value of integrability breaking parameter to see Wigner-Dyson statistics in level spacing. This critical value of the integrability parameter approaches zero in the thermodynamic limit  $N\to\infty$ ~\cite{PhysRevB.90.075152,modak2014finite}.  It has been argued in Ref.~\cite{PhysRevB.99.104205} that at the MBL transition point, the $r$ parameter value is expected to be close to $0.45$ for $N=12$; we have used that as a diagnostic tool to identify the transition point for different values of $V$ (see Fig.~\ref{fig:LevelSpacing}).

\begin{figure}[H]
    \centering
    \includegraphics[width=0.9\linewidth]{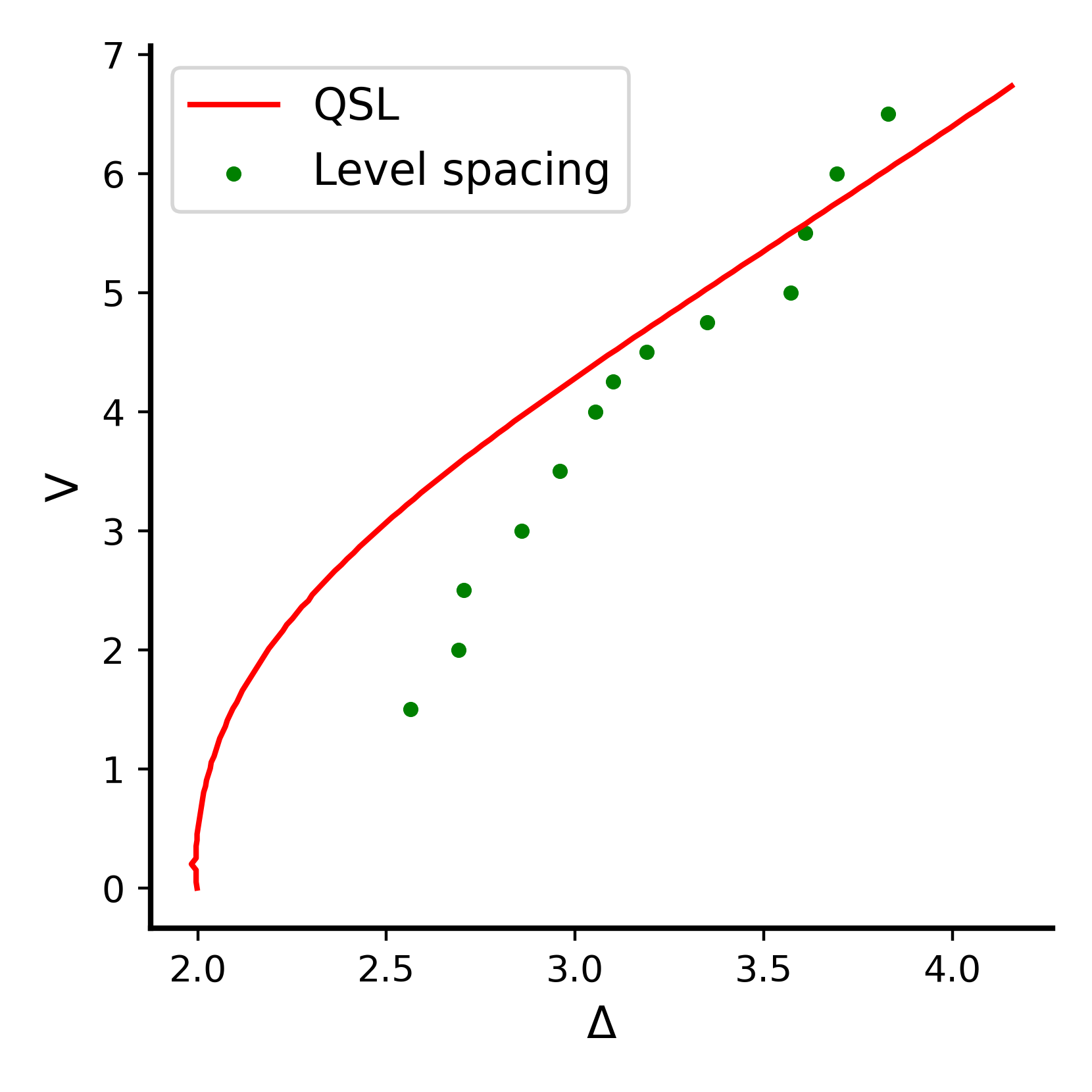}
    \caption{Detection of many body localization transitions from QSL and level spacing characteristics. We have taken lattice site $N = 12$ with half filling .}
    \label{fig:MBLQSL}
\end{figure}

Next, we focus on finding the MBL transition point from QSL. First, we take the initial state of the system as the ground state of the pre-quench Hamiltonian $H_{MBL}(J=1,\Delta_i=0, V)$ and find the QSL as a function of $\Delta$ for the post-quench Hamiltonian $H_{MBL}(J=1,\Delta > 0, V)$. Similarly, we also find QSL for the initial state that is the ground state of the pre-quench Hamiltonian $H_{MBL}(J=1,\Delta_i = 100, V)$ and the post-quench Hamiltonian is $H_{MBL}(J=1,\Delta < \Delta_i, V)$. As discussed in the previous section, the value of $\Delta$ for which these two QSLs will intersect, we identify that $\Delta$ as a transition point for a fixed $V$.  We plot this transition point for different interaction strengths  $V$ in Fig.~\ref{fig:MBLQSL}. We plot transition points obtained by level spacing statistics in the same figure. (Note that we have recognized the transition point by identifying $\Delta$ for which the value of $r$ parameter is $0.45$~\cite{PhysRevB.99.104205}). Remarkably, the transition points obtained from level spacing and QSL are in great agreement for $V\in [4,6]$. However, we like to emphasize that for system size $N\leq16$ (i.e., what is accessible to us using exact diagonalization for interacting system), if  $V$ is reasonably small ($V<2$) or large ($V>6.5$), the maximum value of $r$ parameter remains much smaller than  $0.53$, which makes our $r$ parameter analysis to detect the ergodic to MBL transition extremely inefficient for such parameters (see Appendix. D).
%+++++++++++++++++++END OF SUBSECTION++++++++++++++++++++++++++++++++++

\section{Conclusions}
Our main questions are to identify the dynamical regimes of QSL applicability for the many-body physical model and to study the QSL for a many-body system that undergoes phase transitions. 
To study these questions, we consider the non-interacting Aubry-Andre model, the Wannier Stark model that shows localization transitions, and the interacting many-body model that shows MBL transitions. We find an outstanding agreement between the localization delocalization transition point obtained using QSL and the one we found from the Normalized Participation Ratio (NPR) in the case of Aubry-Andre and Wannier Stark Hamiltonian. Furthermore, we also compare ergodic to Many Body Localization (MBL) transitions found by QSL and level spacing characteristics of the many-body interacting Hamiltonian.

In the case of Aubry-Andre and Wannier Stark Hamiltonian, we find that for the quench across the phase boundary, the dynamics of the Loschmidt echo can be well described by QSLs. We prove, in the small time limit, the MT bound is always tighter compared to ML and ML$^*$ bounds, but for the time of orthogonalization $t_{\perp}$, the tighter bound can also be ML and ML$^*$
depending on the average energy and energy uncertainty of the initial state. We also show analytically for extreme quenches i.e. $\Delta \to 0$ to $\Delta \to \infty$ or vice versa, the MT bound is always tighter for all the states. Moreover, a comparison has been made between the time required for the 1st zero of the Loschmidt echo and the QSL bound corresponding to the time of orthogonalization. While the exact values are a bit different, both of them qualitatively show similar behavior as a function of the quench parameter, e.g., in the case of a quench from delocalized to localized phase, they scale inversely with $\Delta$ and for the opposite quench, they remain independent of $\Delta$.  Most strikingly, we find that if the pre-quench Hamiltonian is either deep inside the delocalized phase or deep into the localized phase, for both cases, the  $\tau_{QSL}$ remains the same if the post-quench Hamiltonian corresponds to the Hamiltonian at the transition point. This fact can be used as a diagnostic tool to detect the transition point. 
Moreover, we have also used the same QSL protocol for interacting systems that show ergodic to MBL transition. First, we use the level spacing statistic to identify the transition point and then compare it with the one obtained using the QSL protocol to find a very good agreement for a certain parameter regime.

The time development of mean return probability of the projectors provides a universal bound~\cite{vikram2024exact,vikram2024proof} for the spectral form factor~\cite{haake1991quantum,muller2004semiclassical,muller2004semiclassical,vikram2023dynamical} through which the minimum time required for certain versions of scrambling can be detected. Also, the universal results on the minimum time required for the scrambling and, thereby, the time at which the application of the equilibrium thermodynamics is studied. The dependency of the QSL on the initial state has been averaged by considering mean return probability, a function of QSL. Our work complements these studies, in which we study localization transitions, and even though QSL depends on the initial state, the quenching from two extreme states (completely localized and completely delocalized states) provides a kind of averaging, and QSL captures the transition and dynamics effectively. We expect that QSL will play an important part in further understanding of many-body theory and phase transitions, especially it will be interesting to study the role of the universal bound proposed in Ref.~\cite{vikram2024exact,vikram2024proof} in future studies.

While there exists a large number of tools to detect localization-delocalization transition points such as participation ratio~\cite{chatterjee2023one,modak2016integrals}, entanglement entropy~\cite{modak2018criterion}, energy spacing statistics~\cite{chatterjee2023one}, observational entropy~\cite{modak2022observational},  and so on, computation-wise $\tau_{QSL}$ is much simpler compared to any of them. It does not involve diagonalization of the entire Hamiltonian, nor does one need to compute the exact dynamics, one simply needs to calculate energy expectation and the energy variance. Given the Hamiltonian of our interests,  are usually sparse matrices, such operations are much less cumbersome. We believe, especially, in the case of the many-body Hamiltonian, to detect ergodic to many body localization transition, our diagnostic tool should be a great advantage over the existing ones.
%\textcolor{red}{ We believe, especially, in the case of the many-body Hamiltonian, to detect ergodic to many-body-localization transition~\cite{abanin2019colloquium}, our diagnostic should be a great advantage over the existing ones, which we plan to investigate in our future studies.}

 \section{ACKNOWLEDGEMENTS}
RM acknowledges the DST-Inspire fellowship by the
Department of Science and Technology, Government of
India, SERB start-up grant (SRG/2021/002152). SA
acknowledges the start-up research grant from SERB,
Department of Science and Technology, Govt. of India
(SRG/2022/000467).

\appendix
\section{Details of the calculation of $\langle H^2(\Delta_f=0) \rangle$}
In this section, we discuss the details of the calculation $\langle H\rangle$ and $\langle H^2 \rangle$ for extreme quench i.e. $\Delta_i\to \infty$ to  $\Delta_f\to 0$. 
At $\Delta_f=0$ diagonal entries of the Hamiltonian are zero. Thus, the Hamiltonian and square of the Hamiltonian take form,
\begin{equation*}
    H = -J\sum_{k=1}^N(c_{k+1}^\dagger c_k + c_k^\dagger c_{k+1}),
\end{equation*}

\begin{equation*}
    H^2 = J^2\sum_{k=1}^N\sum_{l=1}^N(c_{k+1}^\dagger c_k + c_k^\dagger c_{k+1})(c_{l+1}^\dagger c_l + c_l^\dagger c_{l+1})
\end{equation*}

\begin{equation*}
    \begin{aligned}
        H^2 =& J^2\sum_{k=1}^N\sum_{l=1}^N(c_{k+1}^\dagger c_{k}c_{l+1}^\dagger c_{l} + c_{k}^\dagger c_{k+1}c_{l+1}^\dagger c_{l}\\
        &+ c_{k+1}^\dagger c_{k}c_{l}^\dagger c_{l+1} + c_{k}^\dagger c_{k+1}c_{l}^\dagger c_{l+1})
    \end{aligned}
\end{equation*}
\newline

As the initial state is taken as one of the eigenstates of the Hamiltonian $H(\Delta_i\rightarrow\infty)$,  they have the form $|\psi\rangle=c_m^\dagger|0\rangle$; $m=1,2....,N$. The expectation value of the Hamiltonian with respect to the above initial states can be computed as,

\begin{equation*}
    \langle H\rangle = \sum_{k=1}^N[ \langle0|c_mc_{k+1}^\dagger c_kc_m^\dagger|0\rangle + \langle0|c_mc_{k}^\dagger c_{k+1}c_m^\dagger|0\rangle],
\end{equation*}

\[\langle H \rangle = 0\]
\vspace{1pt}
Similarly, the average of the square can be written as,
\vspace{1pt}
\begin{equation}
    \begin{aligned}
        \langle H^2\rangle =& J^2 \sum_{k=1}^N\sum_{l=1}^N[
    \langle0|c_mc_{k+1}^\dagger c_{k}c_{l+1}^\dagger c_{l}c_m^\dagger|0\rangle\\
    &+ \langle0|c_mc_{k}^\dagger c_{k+1}c_{l+1}^\dagger c_{l}c_m^\dagger|0\rangle\\
    &+ \langle0|c_mc_{k+1}^\dagger c_{k}c_{l}^\dagger c_{l+1}c_m^\dagger|0\rangle\\
    &+ \langle0|c_mc_{k}^\dagger c_{k+1}c_{l}^\dagger c_{l+1}c_m^\dagger|0\rangle].
    \end{aligned}
    \label{LongOne}
\end{equation}\\

For an initial state where $m \neq 1$ or $m \neq N$, the first term and last term go to zero and the other term equals one. Hence $\langle H^2\rangle = 2J^2$

For the initial state where $m=1$ or $m=N$, the third or fourth term along with the first and second term goes to zero. Hence for such a state, $\langle H^2\rangle = J^2$

\section{Participation ratio calculations to detect transition point}
In this section, we use the participation ratio as another diagnostic tool to detect the localization-delocalization transition point. 
The Normalized Participation Ratio (NPR)~\cite{PhysRevB.105.214203} of an eigenstate $|\psi_k\rangle$ of the Hamiltonian $H(\Delta)$ (we took $J=1$), is given by%~\cite{chatterjee2023one,modak2016integrals},

\begin{equation*}
    NPR_k = \frac{1}{N\sum_{n=1}^N|\langle n|\psi_k\rangle|^4},
\end{equation*}

Here $|n\rangle = c_n^\dagger|0\rangle$, stands for the Fock space basis for the lattice. If states are completely localized on a given site, then, $NPR = 1/N$. If states are completely delocalized then, $NPR = N$. Here, we calculate the average NPR by taking an average over all the eigenstates of Hamiltonian $H(\Delta)$.

\begin{figure}[H]
    \centering
    \includegraphics[width=0.9\linewidth]{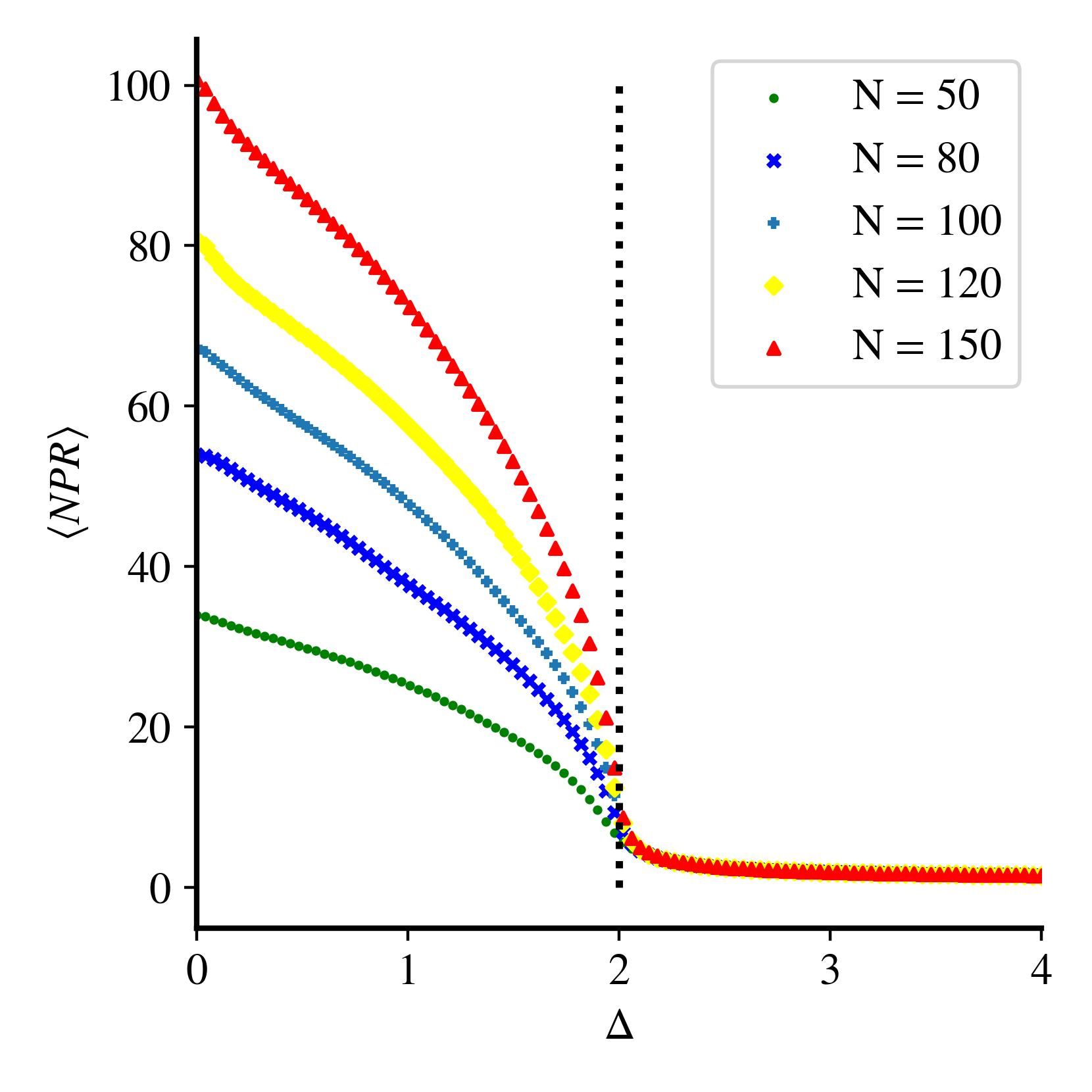}
    \caption{Average normalized participation ratio vs $\Delta$ for incommensurate model. The dotted line indicates phase transition calculated from QSL.}
    \label{fig:Appendix1}
\end{figure}

In the case of the AA model, from figure \ref{fig:Appendix1} it is clear that there is a phase transition point at $\Delta = 2$. In the region $\Delta < 2$, all the states are delocalized (mean NPR increases with $N$), and for $\Delta > 2$ all the states are localized (mean NPR is independent of $N$).

\begin{figure}[H]
    \centering
    \includegraphics[width=0.9\linewidth]{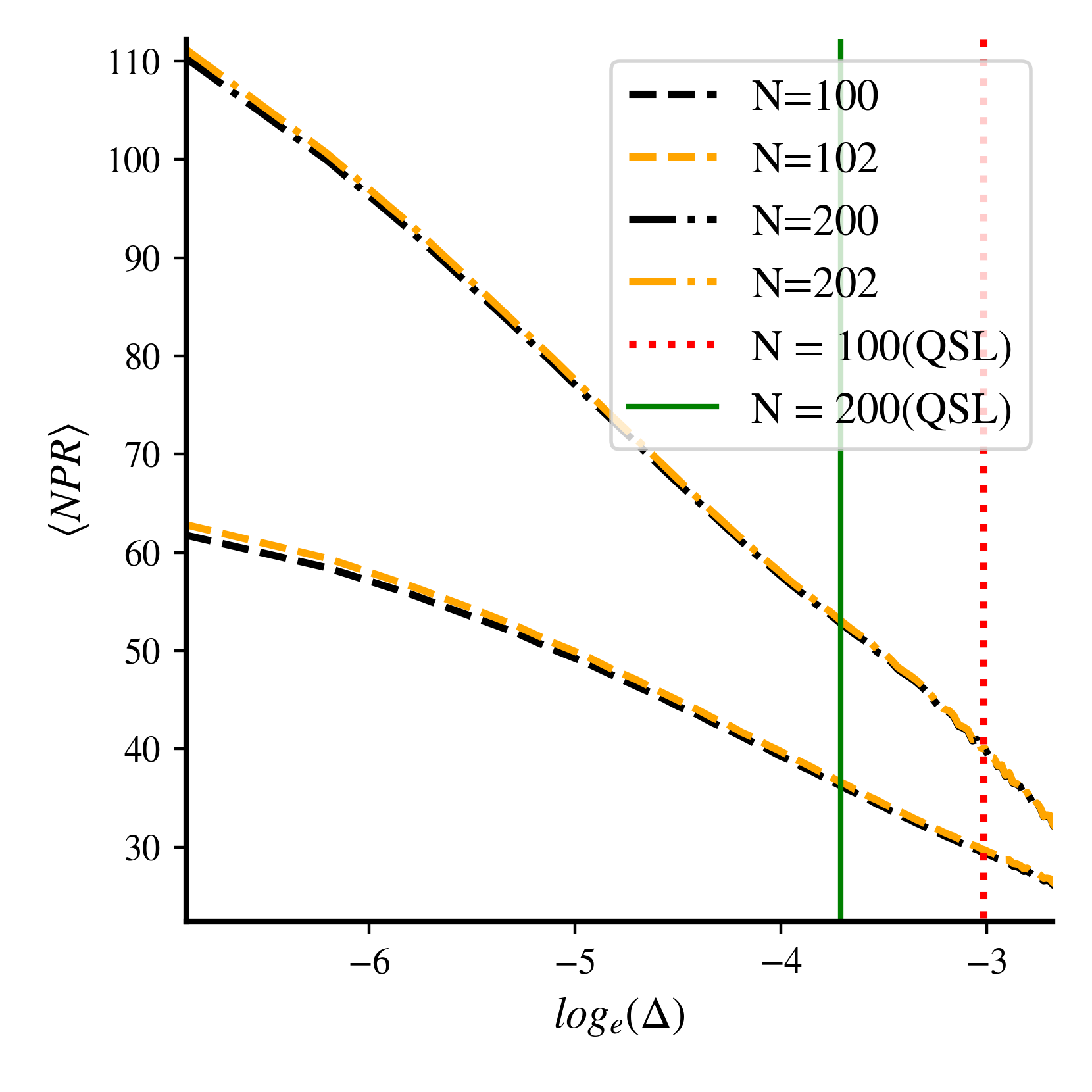}
    \caption{Average normalized participation ratio vs $\Delta$ for Wannier Stark model. The vertical line indicates the phase transition calculated from QSL.}
    \label{fig:Appendix2}
\end{figure}

For the finite-size Wannier Stark model, it is known that the transition point varies with system size. 
However, finding such a $ N$- dependent transition point is quite non-trivial. In the case of the AA model, such an issue does not arise. Hence, we could plot the NPR for different values of $N$, and if $N$ is sufficiently large, we could easily distinguish the delocalized and localized phases.  
In the case of the Wannier Stark model, we expect that if we plot NPR for two system sizes $N$ and $N+\Delta N$ (where $\Delta N$ is small) beyond a certain $\Delta_c$, the NPR data will almost become indistinguishable, that can be identified as the transition point for system size $N$. 
To see this, we plot $\langle NPR \rangle$ for two systems whose system size differences are very small. In figure \ref{fig:Appendix2}, we consider system sizes of $100$ and $102$; the point from which $\langle NPR \rangle$ of both systems start overlapping will be the transition point. We also plot the same for system sizes $200$ and $202$. From the plot, it is clear that the transition point tends to decrease with increasing system size. The vertical lines in the figure, which represent QSL, also predict the same. Furthermore, the QSL prediction of the transition points is approximately equal to those points from which the data of NPR start overlapping. For $N \to \infty$, NPR data will start overlapping even from an infinitesimal value of $\Delta$, and so will the transition point obtained from the QSL, implying any tiny value of potential $\Delta$ is enough to localize all the eigenstates.
%\textcolor{green}{For the Stark model it is known that any finite value, $\Delta > 0$ is enough for localizing system. We demonstrate it with NPR plot in figure \ref{fig:Appendix2}. It can be seen that as system size increases, transition point tends to zero, indicating that for a thermodynamic limit any finite $\Delta$ is enough to localize system. In the figure, QSL does not coincide with steep fall of the curve but it follows the same pattern, i.e. phase transition point tend to zero as system size increases. Thus it can be deduced that at the thermodynamic limit, from QSL, for any $\Delta > 0$ the system localizes.}

\section{Finite size dependence on QSL of AA model}

For the thermodynamic limit, in equation \eqref{Average} and \eqref{AverageSquare}, we have approximated averages of $\cos$ and $\cos^2$ as $0$ and $1/2$, respectively. But for a finite value of the $N$, we can calculate the exact summation as follows.

\begin{equation}
    \sum_{n=1}^N \cos (2\pi\alpha n) = \sum_{n=1}^N \left( \frac{e^{2i\pi\alpha n} + e^{-2i\pi\alpha n}}{2}  \right)
    \label{temp1}
\end{equation}
From geometric series, it can be proved that,
\begin{equation}
    \sum_{n=1}^N x^n = \frac{x(1-x^n)}{1-x}
\end{equation}
Using this, equation \eqref{temp1} can be simplified to,
\begin{equation}
\begin{aligned}
    & g(N,\alpha) \equiv \sum_{n=1}^N \cos (2\pi\alpha n) = \\
    & \frac{\cos(2\pi\alpha) - \cos(2\pi\alpha(N+1)) + \cos(2\pi\alpha N) - 1}{2(1-\cos(2\pi\alpha))}
\end{aligned}
\end{equation}

\noindent Similarly, it is easier to obtain

\begin{equation}
    \sum_{n=1}^N \cos^2 (2\pi\alpha n) = \frac{N}{2} + g(N,2\alpha)
\end{equation}

\noindent Thus, uncertainty reads as (referring to equation \eqref{Average} \& \eqref{AverageSquare})

\begin{equation}
    \Delta H = \Delta_f \sqrt{\left( \frac{1}{2} + \frac{1}{N}g(N,2\alpha) - \frac{1}{N^2}g^2(N,\alpha) \right)}
\end{equation}

Thus, at the thermodynamic limit, leading order term of $(\Delta H)^2$ for AA model is $\Delta^2_f / 2$. In case of the Wannier Stark model, we observed that the leading order of $(\Delta H)^2$ itself was proportional to $\Delta_f^2 (N^2 - 1)$, with a missing higher order terms.

\section{Level spacing statistics for $V=0.5$ and $V=10$}
\begin{figure}[H]
    \centering
    \includegraphics[width=0.9\linewidth]{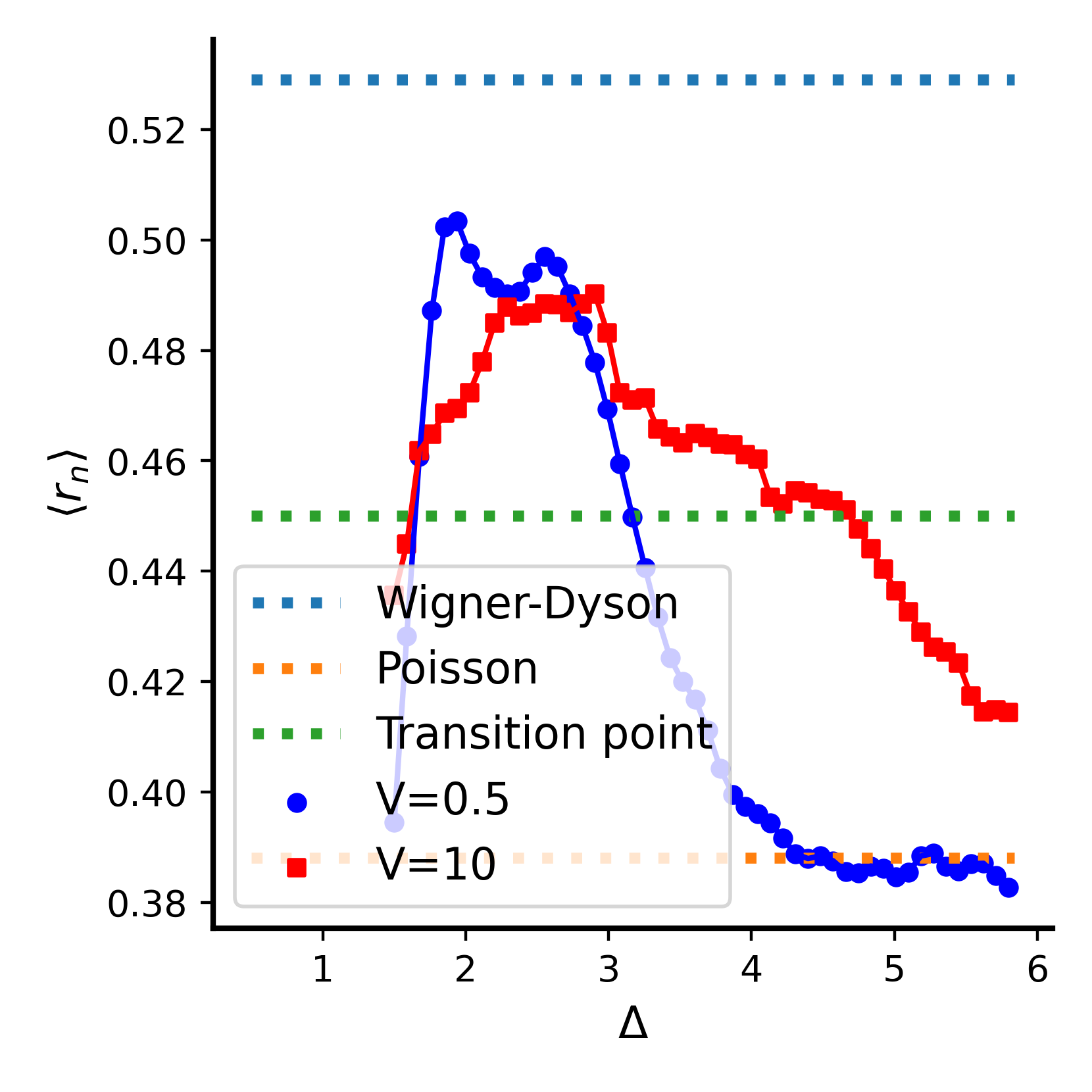}
    \caption{Plot of average level spacing statistics vs $\Delta$ for different interaction potential $V$. System size is taken to be $L=12$ with half filling.}
    \label{fig:LevelSpacingOutside}
\end{figure}

As we discussed in section IV D in the main text, for small $V$ and large $V$, the maximum value of the $r$ parameter remains much smaller than $0.529$. This is illustrated in the figure \ref{fig:LevelSpacingOutside}. This makes our level-spacing analysis inefficient in detecting the transition point. 

%***************BIBILOGRAPHY*************************
\bibliography{ref}
%***********END OF BIBILOGRAPHY**********************
\end{document}